\documentclass[aip,reprint]{revtex4-1}

\usepackage{graphicx}
\usepackage{dcolumn}
\usepackage{bm}

\draft 

\begin{document}


\title{Tapered fiber coupling of single photons emitted by a deterministically positioned single nitrogen vacancy center} 



\author{Lars Liebermeister},
\email[]{email: lars.liebermeister@physik.uni-muenchen.de} 
\author{Fabian Petersen},
\author{Asmus v. M\"unchow},
\author{Daniel Burchardt},
\author{Juliane Hermelbracht},
\author{Toshiyuki Tashima},

\affiliation{Fakult\"at f\"ur Physik, Ludwig-Maximilians-Universit\"at
  M\"unchen, 80799 M\"unchen Germany}
\author{Andreas W. Schell},
\author{Oliver Benson},
\affiliation{Institut f\"ur Physik, Humboldt-Universit\"at zu Berlin, 12489
  Berlin Germany}
\author{Thomas Meinhardt},
\author{Anke Krueger},
\affiliation{Institut f\"ur Organische Chemie, Universit\"at W\"urzburg, 97074
  W\"urzburg Germany},
\affiliation{Wilhelm Conrad Roentgen Research Center for Complex Materials
  Systems, Universit\"at W\"urzburg, 97074 Wuerzburg Germany},
\author{Ariane Stiebeiner},
\author{Arno Rauschenbeutel},
\affiliation{Atominstitut, Technische Universit\"at Wien, 1020 Wien Austria}
\author{Harald Weinfurter},
\author{Markus Weber}
\email[]{email: markusweber@lmu.de} 
\affiliation{Fakult\"at f\"ur Physik, Ludwig-Maximilians-Universit\"at
  M\"unchen, 80799 M\"unchen Germany},
\affiliation{Max-Planck-Institut f\"ur Quantenoptik, 85748 Garching, Germany}


\date{\today}

\begin{abstract}
A diamond nano-crystal hosting a single nitrogen vacancy (NV) center is
optically selected with a confocal scanning microscope and positioned
deterministically onto the subwavelength-diameter waist of a tapered optical
fiber (TOF) with the help of an atomic force microscope. Based on this
nano-manipulation technique we experimentally demonstrate the evanescent
coupling of single fluorescence photons emitted by a single NV-center to the
guided mode of the TOF. By comparing photon count rates of the fiber-guided
and the free-space modes and with the help of numerical FDTD simulations we
determine a lower and upper bound for the coupling efficiency of
$(9.5\pm0.6)\%$ and $(10.4\pm0.7)\%$, respectively. Our results are a
promising starting point for future integration of single photon sources into
photonic quantum networks and applications in quantum information science.
\end{abstract}


\maketitle 

Efficient collection of single photons radiated by a single solid state
quantum emitter -- like the nitrogen vacancy (NV) center in diamond
\cite{Kurtsiefer00} -- is an important prerequisite for future applications in
applied physical and quantum information science, like ultra-sensitive
fluorescence spectroscopy and linear optical quantum computation
\cite{KLM01,Kok07,Jennewein11}. A standard technique for fluorescence
collection is confocal microscopy. However, when applied to defect centers in
bulk diamond, total internal reflection limits the collection efficiency to
few percent. Recently, the collection efficiency of NV-fluorescence has been
increased by one order of magnitude by combining confocal microscopy with
solid immersion lenses (SILs) \cite{Hadden10,Wrachtrup10,Marseglia11},
respectively photonic nanowires \cite{Babinec10}. In the latter system the
improvement is based on efficient coupling of NV-fluorescence photons to the
strongly confined mode (HE$_{11}$) \cite{Domokos02,Fu08,Loncar09} of diamond
nanowires. For defect centers in diamond nano-crystals, tapered optical fibers
(TOFs) \cite{Stiebeiner10} with a subwavelength diameter waist are a
particularly attractive alternative platform. Due to the strong evanescent
field at the surface, such TOFs promise coupling efficiencies up to $36\%$
\cite{Klimov04,Kien05} and approaching unity when combined with Bragg-grating
cavities \cite{Kien09,Wuttke12}.

\begin{figure}
\includegraphics[width=8cm]{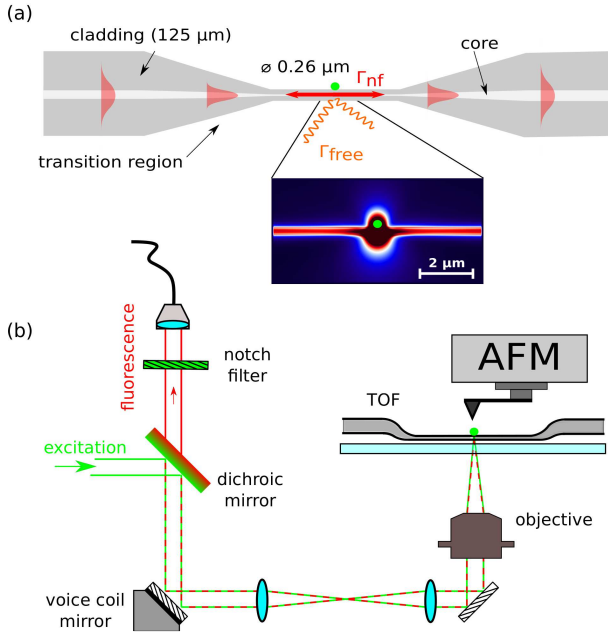}
\caption{(a) Schematic of evanescent coupling of single photons emitted by a
single nitrogen vacancy center (hosted in a diamond nano-crystal) to a single
guided mode of a tapered optical fiber (TOF). Inset: FDTD simulation of the
intensity distribution of a radiating NV-center coupled to the nano-fiber. (b)
Schematic diagram of the experimental setup. An inverted optical confocal
scanning microscope is used to analyze the fluorescence properties of
individual diamond nano-crystals. In addition, an atomic force microscope (AFM)
is employed for in-situ nano-manipulation of individual crystals. Operating
both microscopes at the same time allows to position a diamond nano-crystal
(hosting only a single NV-center) on demand onto the apex of the optical
nano-fiber.} \label{setup_fiber:fig}
\end{figure}

Until now, evanescent coupling of fluorescence photons to a single guided mode
of a TOF has been achieved for various solid state quantum emitters
\cite{Fujiwara11,Schroeder12,Yalla12a,Yalla12b}, molecules
\cite{Stiebeiner09}, and laser-cooled atomic vapors \cite{Nayak07}. To bring
these emitters into the strong evanescent optical field at the surface of the
nano-fiber several non-deterministic deposition techniques like dip-coating
\cite{Fujiwara11,Schroeder12}, picoliter-dispensers \cite{Yalla12a,Yalla12b},
and optical surface traps \cite{Vetsch10} have been applied. However, for real
applications in quantum information science, e. g., the photonic quantum-bus
mediated coupling of NV-centers in a lattice \cite{Ciccarello10},
deterministic positioning of single solid state quantum emitters onto the
submicron waist of a TOF with nm position control is desirable. In this letter
we demonstrate significant steps towards deterministic coupling of a single
solid state quantum emitter to a tapered optical fiber, i.e. (i) the on-demand
positioning of a single diamond nano-crystal hosting a single NV-center onto
the nanofiber-waist and (ii) the evanescent coupling of single fluorescence
photons to a single guided mode of the TOF (see Fig. \ref{setup_fiber:fig},
a).

Our tapered optical fiber is produced from a standard optical single mode
fiber, drawn
down to a waist diameter of 260 nm with a fiber-pulling-rig
\cite{Warken07,Stiebeiner10}. Due to the subwavelength diameter of the waist,
an NV-center close to the surface of the nanofiber experiences a strong
optical field \cite{Kien04} of the fundamental guided mode HE$_{11}$. This
results in a small effective mode area ${\cal A}=[\int_A
\epsilon(\vec{r})|\vec{E}(\vec{r})|^2\,\text{d}^2\vec{r}\,]/
[\epsilon(\vec{r}_i)|\vec{E}(\vec{r}_i)|^2\,]$. Here $\epsilon(\vec{r})$ is
the electric permittivity and $\vec{E}(\vec{r})$ the electric field at a
position $\vec{r}$, whereas $\epsilon(\vec{r}_i)$ and $\vec{E}(\vec{r}_i)$ are
the permittivity and the field at the position $\vec{r}_i$ of the NV
center. As the spontaneous emission rate into the nanofiber is given by
\cite{Domokos02} $\Gamma_{nf}=\sigma_A \Gamma_0 / (2\cal A)$ and as $\cal{A}
\approx$ $\lambda^{2}$ this results in strong coupling of NV-emission to the
nanofiber. Here $\sigma_A=3\lambda^2/(2\pi)$ is the radiative atomic cross
section, $\lambda$ the wavelength of the emitted photons, and $\Gamma_0$ the
spontaneous decay rate into free space modes in absence of the nanofiber. To
obtain a more quantitative prediction for our coupling efficiency we performed
numerical FDTD simulations with MEEP \cite{MEEP}. Depending on the orientation
of a linearly polarized point dipole (situated 10 nm above the nano-fiber
surface) we get maximum coupling efficiencies of
$27.5\%, 15.6\%,$ and $34.9\%$ for tangential, parallel, and radial
polarization, respectively.

\begin{figure}
\includegraphics[width=8cm]{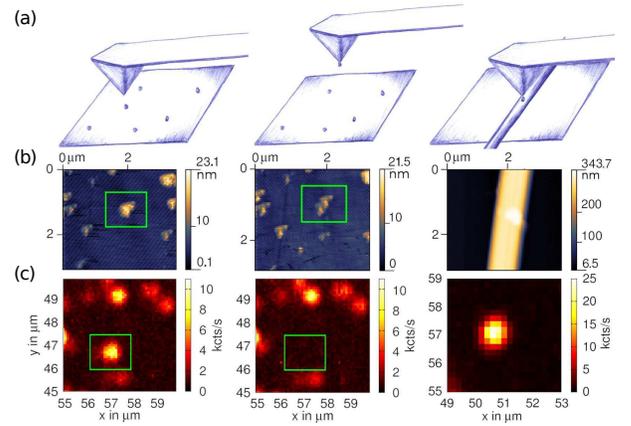}
\caption{(a) Schematics of the on-demand positioning of single fluorescing
diamond nano-crystals onto the apex of an optical nano-fiber. (b) Topography
and (c) optical scan before pick-up, after pick-up, and after placing onto the
optical fiber (from left to right).} \label{pick_place:fig}
\end{figure}

On-demand highly accurate picking and placing of single diamond nano-crystals
is achieved using a recently introduced nano-manipulation technique
\cite{Schell11}, based on a confocal fluorescence microscope combined with a
commercial atomic force microscope (AFM) (see Fig. \ref{setup_fiber:fig},
b). With this device we simultaneously monitor the topography (see
Fig. \ref{pick_place:fig}, b) and the respective optical response (see
Fig. \ref{pick_place:fig}, c) of NV-centers in diamond nano-crystals before,
during, and after assembly (see Fig. \ref{pick_place:fig}, from left to
right). First, diamond nano-crystals \cite{Krueger} (mean diameter $\approx
25$ nm) hosting single NV-centers are identified on a fused silica substrate
by observing photon anti-bunching of the emitted fluorescence light. Second,
the selected nano-diamond is picked up with the AFM. Third, the TOF is
deposited on a clean substrate and the nano-crystal is placed onto the
nano-fiber waist. Finally, the TOF is detached from the fused silica
substrate.

For all optical investigations, the NV-center on the nano-fiber is excited via
the confocal microscope with a continuous wave laser at a wavelength of 532 nm
(perpendicular to the TOF). NV-fluorescence is collected confocally with the
microscope objective (nominal numerical aperture NA=0.75) and via the tapered
optical fiber. The spectrum of the collected fluorescence photons is recorded
with a CCD spectrometer with 1 nm resolution. In addition, the photon
statistics of the fluorescence light is analyzed via the second-order
correlation function $g^{(2)}(\tau)$, measured in a Hanbury-Brown-Twiss (HBT)
configuration. Two options are possible: either with two single-photon
detectors (SPD) following a bulk beam splitter at the output fiber of the
confocal microscope \cite{Kurtsiefer00} or with two SPDs positioned at the
ends of the TOF. Here we emphasize that in the latter setting the fiber acts
as intrinsic beam splitter and once photon anti-bunching is observed it
verifies the coupling of single photons to the guided mode of the
TOF. Differences $\tau$ of detection times of photon pair events are recorded
with a time-stamp-unit with $77$ ps time resolution and stored in a histogram
with a time bin width of $t_{bin}=0.924$ ns. To obtain $g^{(2)}(\tau)$ from
this delay time histogram, we divide the number of entries in each time bin by
its average value for long detection time differences $\tau=0.7$ $...$ $1.1
\mu$s.

\begin{figure}
\includegraphics[width=8cm]{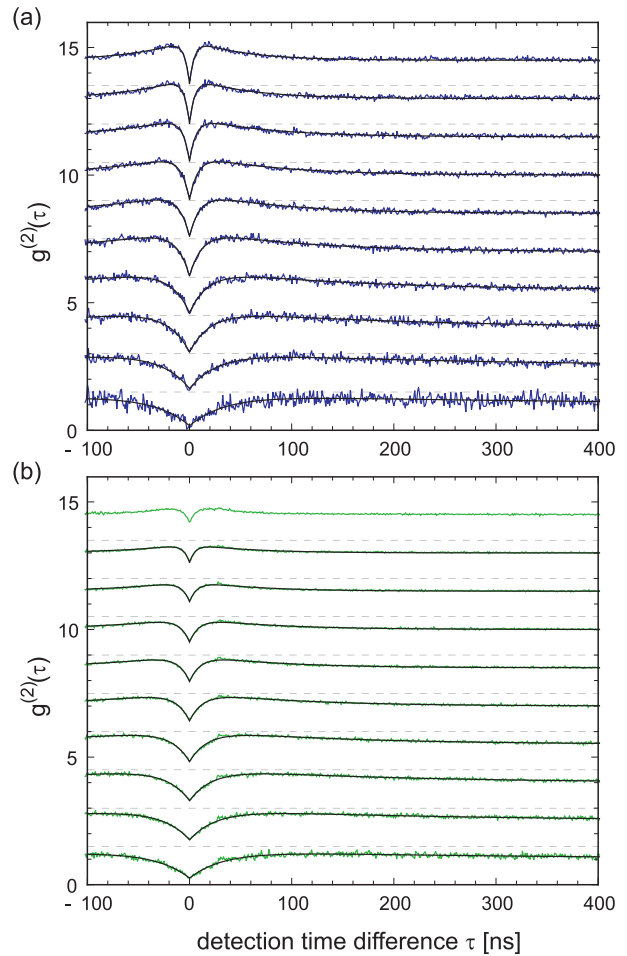}
\caption{Second-order correlation function $g^{(2)}(\tau)$ of fluorescence
light emitted by a single NV-center for different excitation powers
$P=(0.5,1.0,1.5,2,3,4,5,6,8,10)$ mW (from bottom to top), coupled to an
optical nano-fiber of 260 nm diameter. The NV-center is excited perpendicular
to the optical fiber via a confocal microscope. Fluorescence photons are
collected (a) via the objective of the confocal microscope and (b) via
coupling to the guided mode of the optical nano-fiber. The anti-bunching dip
at zero detection time difference $\tau=0$ demonstrates the non-classical
character of the fluorescence light as well as the coupling of the NV-center
emission to the optical nano-fiber. For better discrimination, the curves are
shifted vertically by increments of 1.5} \label{HBT:fig} 
\end{figure}

\begin{figure}
\includegraphics[width=8cm]{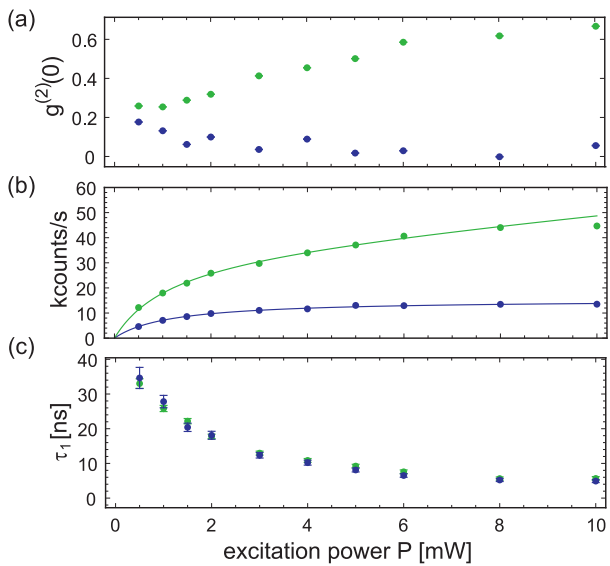}
\caption{(a) Second-order correlation function $g^{(2)}(0)$ at detection time
difference $\tau=0$, (b) detected count rate, and (c) fit-parameter $\tau_1$
as a function of the excitation power $P$. Blue data points stem from confocal
collection whereas green data points from nano-fiber collection,
respectively.} \label{HBTfitparameter:fig}
\end{figure}

To analyze the evanescent coupling of fluorescence light emitted by a
single NV-center to the guided mode of the TOF, we measured $g^{(2)}(\tau)$
for different excitation powers $P$ (see Fig. \ref{HBT:fig}, b). For
comparison the second-order correlation function is recorded simultaneously
via confocal collection (see Fig. \ref{HBT:fig}, a). Both sets of measured
$g^{(2)}(\tau)$ functions are then fitted with the modified three-level model
\cite{Kurtsiefer00,Wang07}
\begin{equation}
g^{(2)}(\tau) = 1 + p_f^2[c
e^{-\frac{|\tau|}{\tau_1}}-(1+c)e^{-\frac{|\tau|}{\tau_2}}],
\end{equation} 
which neglects intensity dependent deshelving of the meta-stable state. The
parameter $p_f$ is the probability that a detected photon event stems from a
single NV-center. $\tau_1$ describes the pump-power dependent slope of the
anti-bunching dip, $\tau_2$ governs the decay of $g^{(2)}(\tau)>1$ for
intermediate detection time differences, while $c$ determines its
amplitude. To directly quantify the influence of fluorescence which does not
stem from the NV-center on the quality of our non-classical single-photon
source we plot the fitted value of the second-order correlation function for
zero detection time difference, i.e. $g^{(2)}(0)=1-p_f^2$ (see
Fig. \ref{HBTfitparameter:fig}, a). For low excitation powers $g^{(2)}(0)$ is
determined by the dark count rates of the SPDs and residual background
fluorescence. Increasing $P$, in confocal collection the influence of the dark
count rate and the residual background fluorescence decreases, leading to
almost perfect anti-bunching. In fiber-based collection, we observe an
excitation power dependent increase of $g^{(2)}(0)$ from $0.26$ to
$0.67$. This points to additional uncorrelated photons that are mainly caused
by intrinsic fluorescence of the fiber core, generated by Rayleigh-scattered
excitation laser light that is coupled into the nano-fiber.

As a next step, we evaluate the nano-fiber coupling efficiency
$\beta=\Gamma_{nf}/(\Gamma_{nf}+\Gamma_{free})$, which is defined as the ratio
of the radiative decay-rate into the nano-fiber $\Gamma_{nf}$ over the total
radiative decay rate $\Gamma_{rad}=\Gamma_{nf}+\Gamma_{free}$. Here,
$\Gamma_{free}$ is the decay rate into free-space modes in the presence of the
nano-fiber. First, to determine $\Gamma_{free}$, we fit the power dependent
confocal count rate (blue data points in Fig. \ref{HBTfitparameter:fig}, b)
with the model $kP/(P+P_{sat})$, where $k$ is the count rate for excitation
power $P\rightarrow \infty$. From this least square fit, we get a saturation
power of $P_{sat}=1.17$ mW and a free-space saturation count rate of
$C_{free}=7.70 \times 10^3$ s$^{-1}$. The corresponding decay rate
$\Gamma_{free}$ into free-space modes can be determined from $C_{free}$ by
taking into account the fraction of photons collected by the microscope
objective (effective numerical aperture NA$^*=0.32\pm0.01$), transmission
losses from the focal spot to the single photon detectors (SPD), and the
quantum efficiency $\eta=0.65$ of the SPDs. With the help of numerical FDTD
simulations \cite{supplement} we get a lower and upper bound for
$\Gamma_{free}$ of $(1.7\pm0.1) \times 10^6$ s$^{-1}$ and $(1.8\pm0.1) \times
10^6$ s$^{-1}$, respectively.

Next, we calculate how many single photons are scattered into the nano-fiber
per second, i.e. the radiative decay rate $\Gamma_{nf}$. To determine the
saturation count rate in nano-fiber collection, we fit the corresponding power
dependent count rate (green data points in Fig. \ref{HBTfitparameter:fig}, b)
with the model $k^{'}P/(P+P_{sat})+mP$, taking into account an additional
linear increase of the fiber background fluorescence. For this fit, $P_{sat}$
is fixed to the value of the first fit (confocal detection and no fiber
background fluorescence), resulting in a nano-fiber saturation count rate of
$C_{nf}=19.6 \times 10^3$ s$^{-1}$. A simple analysis shows that
$\Gamma_{nf}=C_{nf}/(\eta \sqrt{T_{ges}})$, where $T_{ges}=(2.41\pm0.03)\%$ is
the overall transmission of the TOF-system (includes the intrinsic
transmission of the TOF, transmission from spectral filters, and coupling
losses from the nano-fiber into detector fibers) and $\eta=0.65$ is the
quantum efficiency of the used SPDs. These values lead to
$\Gamma_{nf}=(1.94\pm0.02)\times 10^5$ s$^{-1}$, resulting in a lower and
upper bound for the nano-fiber coupling efficiency $\beta$ of $(9.5\pm0.6)\%$
and $(10.4\pm0.7)\%$, respectively.

This value should be compared with the expected coupling efficiency of a
radiating NV-center (emission wavelength $\lambda=666$ nm, located 10 nm above
the nano-fiber) coupled to a nano-fiber with 260 nm diameter. From
polarization dependent excitation measurements and numerical FDTD simulations
\cite{supplement} we can estimate a lower and upper bound for $\beta$,
yielding $(28.78\pm0.03)\%$ and $(29.22\pm0.03)\%$, respectively. Reduction of
this value by few percent is expected due to the broadband emission spectrum
of the NV-center, however can not explain the discrepancy to our experimental
finding. To clarify this contradiction, also supported by a recent
experimental work with a single CdSe/ZnS nano-crystal \cite{Fujiwara11}, we
plan further experimental and theoretical investigations.

Concluding, as the total decay rate $\Gamma_{tot}=\Gamma_{rad}+\Gamma_{nrad}$,
i.e. the sum of the radiative and nonradiative rate, can be determined from
the measured second-order correlation functions for different excitation
powers \cite{Kurtsiefer00}, we furthermore can give an estimate of the
internal quantum efficiency $QE=\Gamma_{rad}/\Gamma_{tot}$ of the
NV-center. For small excitation powers the fit parameter $\tau_1$ reaches
$\tau_{tot}=1/\Gamma_{tot}= (63\pm9)$ ns (see Fig. \ref{HBTfitparameter:fig},
c), resulting in an upper and lower bound for the quantum efficiency $QE$ of
$(12.9\pm2.0)\%$ and $(11.8\pm1.8)\%$. This finding is in good agreement with
recent QE-measurements from NV-centers in diamond nano-crystals of different
size \cite{Mohtashami12}.

In this work we have demonstrated the on-demand positioning of a single
diamond nano-crystal hosting a single NV-center onto the nanofiber-waist of a
tapered optical fiber and its efficient optical coupling to a guided
nano-fiber mode. The observed coupling efficiency is a promising starting
point for future applications in ultra-sensitive phase
\cite{Kurtsiefer09,Sandoghdar11,Sandoghdar12}, absorption \cite{Kurtsiefer08},
and fluorescence spectroscopy. As a single quantum emitter can shift the phase
of a propagating laser beam by several degrees, this level of nonlinearity
would, e.g., be sufficient to provide a useful photon-photon interaction for
optical quantum information science \cite{Munro05,Hwang11}. Additionally,
replacing the NV-center by a narrowband solid-state emitter like the
SiV-center in diamond \cite{Wang06,Neu11,Neu13} and combining the nano-fiber
with a Bragg-grating cavity \cite{Wuttke12} will pave the way towards the
realization of an efficient single photon source at room temperature, an
essential building block of photonic quantum computing
\cite{KLM01,Kok07,Jennewein11}.

\begin{acknowledgments}
We acknowledge stimulating discussions with David Hunger and funding by the DFG
through the projects FOR1493 (Diamond Materials for Quantum Application) and
the excellence cluster NIM, respectively, and the BMBF through the project
EPHQUAM.
\end{acknowledgments}


\end{document}